\documentclass[12pt,preprint,english]{aastex}
\usepackage{subfigure}
\usepackage{graphicx}

\newcommand{\unit}[1]{\,{\rm #1}}
\newcommand{\Rm}{Re_{\rm m}}
\providecommand{\boldsymbol}[1]{\mbox{\boldmath $#1$}}

\providecommand{\tabularnewline}{\\}

\usepackage{babel}
\shorttitle{2D MRI simulations}
\shortauthors{W. Liu \emph{et al.}}

\begin{document}

\title{Numerical Study of the Magnetorotational Instability in Princeton MRI Experiment}

\author{Wei Liu\footnote{current address: Theoretical Division, Los Alamos National Laboratory, Los Alamos, NM, USA 87545}}

\affil{Center for Magnetic Self-Organization in Laboratory and Astrophysical Plasma, Princeton Plasma Physics Laboratory, Princeton, NJ, USA 08543}
\email{wliu@lanl.gov}

%\author{Jeremy Goodman}

%\affil{Princeton University Observatory, Princeton, NJ, USA 08544}

%\author{Hantao Ji}

%\affil{Center for Magnetic Self-Organization in Laboratory and Astrophysical Plasma, Princeton Plasma Physics Laboratory, Princeton University, P.O. Box
%451, Princeton, NJ, USA 08543 }

\begin{abstract}
  In preparation for an experimental study of magnetorotational
  instability (MRI) in liquid metal, we present non-ideal
  axisymmetric magnetohydrodynamic simulations of the nonlinear
  evolution of MRI in the experimental geometry. The simulations adopt
  fully insulating boundary conditions. No-slip
  conditions are imposed at all boundaries. A clear linear phase is observed with reduced linear growth rate. MRI results in an inflowing ``jet" near the midplane and enhances the angular momentum transport at saturation.

\end{abstract}
\keywords{accretion, accretion disks --- instabilities --- methods: numerical --- MHD}

\maketitle

\section{Introduction}
The magnetorotational instability (MRI) is probably the main cause of
turbulence and accretion in sufficiently ionized astrophysical disks \citep{bh98}, which has 
inspired searches for MRI in Taylor-Couette flow. 
Experiments on magnetized Couette flow aiming to observe MRI have been performed \citep{sisan04,sgg06}. Those experiments have demonstrated MRI-like modes, but not yet for background flows that approximate Keplerian disks.  \citet{sisan04} observed nonaxisymmetric modes and enhanced transport in a spherical experiment, but with a hydrodynamically linearly unstable background state.  \citet{sgg06,rhsgg06,sgg07} demonstrated unsteady behavior in a very low-Rm experiment with a strong toroidal as well as axial field; the relationship between the experimental results and linear instabilities is still controversial \citep{lghj06,pgg07,lgj07,rh07,sj07,lw08}, but if these are indeed the linear modes, then the corresponding ratio of toroidal to axial field in a thin (thickness $\Delta h\ll r$) disk would have to be $\sim r/\Delta h$, where $r$ is the radius of the accretion disk.  
 Some other experiments have been proposed or are still under construction \citep{npc02,vils06,jgk01,gj02}.
 
\begin{figure}[!htp]
\begin{center}
\scalebox{0.6}{\includegraphics{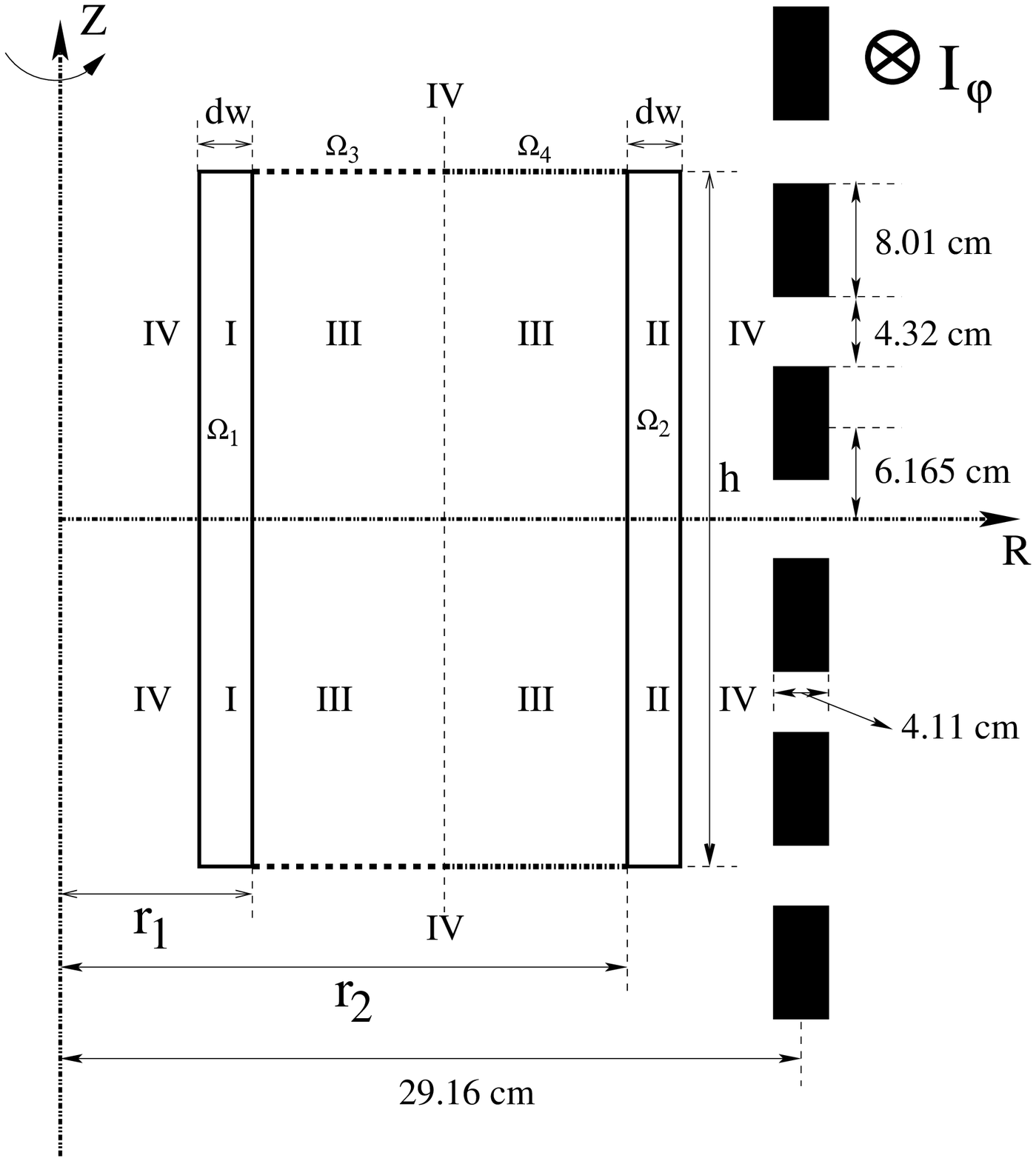}}
\caption{\label{princeton_full_scheme} 
  Computational domain for simulations of Princeton MRI experiment. Region
  (I): Inner steel cylinder, angular velocity $\Omega_{1}$, magnetic resistivity $\eta_{I} $.
  (II): outer steel cylinder, 
  $\Omega_{2}$, $\eta_{II}$. (III): liquid gallium, $\eta_{\rm{Ga}}$; (IV):
  vacuum. Thick dash line: insulating inner ring, $\Omega_{3}$.
  Thick dash-dot line: insulating outer ring, $\Omega_{4}$. 
  Thin dash line: middle plane. Dimensions:
 radius of the inner cylinder,  $r_{1}$; radius of the outer cylinder, $r_{2}$; height of the cylinders, $h$;
  thickness of inner and outer steel cylinder, $d_{w}=0.9525\unit{cm}$; black rectangles, external coils.
  }

\end{center}
\end{figure}

Standard MRI modes will not grow unless both the rotation period and
the Alfv\'en crossing time are shorter than the timescale for magnetic
diffusion.  This requires that both the magnetic Reynolds number
$\Rm\equiv\Omega_{1}r_{1}(r_{2}-r_{1})/\eta_{\rm Ga}$ (see definitions of $\Omega_1$, $\eta_{\rm Ga}$, $r_1$ and $r_2$ in Fig.~\ref{princeton_full_scheme} and their values in Table.~\ref{Ta_parameter}) and the Lundquist
number $S\equiv V_{Az}^{0}(r_{2}-r_{1})/\eta_{\rm Ga}$ be $\gtrsim 1$, where
$V_{Az}^{0}=B_{\infty}/\sqrt{4\pi\rho_{\rm Ga}}$ is the Alfv\'en speed, $\rho_{\rm Ga}$ is the density of liquid Gallium and
$B_{\infty}$ is the background magnetic field parallel to the
angular velocity. Because the magnetic Prandtl number $Pr_{\rm m}\equiv\nu/\eta\sim
10^{-5}-10^{-6}$ in liquid metals, where $\nu_{\rm Ga}$ is the kinematic viscosity of liquid gallium, $Re\gtrsim 10^6$ and fields of several kilogauss for standard MRI
must be achieved in typical experimental geometries. Recent linear analyses have shown that MRI-like mode could grow with much reduced magnetic Reynolds number and Lundquist number in the presence of a helical background magnetic field \citep{hr05,rhss05}, at least for cylinders of infinite or periodic axial extent (Liu et al 2006b). The simulations of this paper,
however, are limited to axial background fields
%% The nature of HMRI and PROMISE is not directly relevant to the present paper, so
%% it is best not to to pick a fight about it here ---JJG.
%Potsdam Rossendorf Magnetic Instability Experiment (PROMISE) group
%claimed to have observed this kind of MRI mode \citep{sgg06}, however
%the wave patterns observed in their experiment turned out to be a
%noise-sustained transient mode rather than a global MRI mode
%\citep{lgj07}.

The experiment is
complicated by the large ($\gtrsim 10^6$) Reynolds number and by
Ekman circulation and Stewartson layers \citep{hf04},
 even though the Princeton MRI
experimental apparatus has been constructed to minimize the circulation by the use of independently controlled split endcaps \citep{kjg04,bsj06,jbsg06}. It is known that
Ekman circulation is significantly modified when the
Elsasser number $\Lambda$  exceeds unity:
$\Lambda=B_{\infty}^{2}/(8\pi\rho_{\rm{Ga}}\eta_{\rm{Ga}}\bar{\Omega})\gtrsim 1$ 
 where $\bar{\Omega}=\sqrt{\Omega_{1}\Omega_{2}}$ is the characteristic rotation frequency
\citep{gp71}.
At 100\% of the maximum designed rotation rate of our experiment
(Table.\ref{Ta_parameter}) and with $I_{\varphi}=1000\;\unit{A}$,
where $I_{\varphi}$ is the external coil current
(Fig.~\ref{princeton_full_scheme}), the Elsasser number $\Lambda=0.36$.
  
There have been few published studies of  the nonlinear saturation of MRI in an experimental Taylor-Couette geometry apart from \citet{kj05,lgj06,umr07a,umr07b}, and even fewer with realistic experimental boundary conditions. As noted in \citet{lgj06}, the nature of saturation of MRI in a Couette flow is essentially different from accretion disks, in which MRI is believed to saturate by turbulent reconnection \citep{fsh00,si01}. In accretion disks, differential rotation arises from a balance between
the gravitational attraction of the accreting body and centrifugal
force. Thermal and turbulent energies are probably small compared to orbital ones, 
so saturation is not likely to occur by modification of the mean flow profile.
In experiments, however, differential rotation is imposed by viscous or other weak 
forces, so that mildly nonlinear MRI might well change the mean rotation profile.

This paper is the second of a series, following \citet{lgj06}. 
%The simulation also mimics an on-going experiment \citep{jgk01,gj02,bsj06,jbsg06,lgj06}. 
Several idealizations made in that paper, notably the vertically periodic
boundary conditions, are dispensed with here.
We discuss the linear phase and nonlinear saturation of MRI in finite
cylinders with realistic fluid and magnetic boundary conditions.
All simulations reported here were performed with the ZEUS-MP
2.0 code \citep{hnf06}, which is a 
time-explicit, compressible, astrophysical ideal MHD parallel 3D code,
to which we have added viscosity, resistivity (with subcycling to reduce
the cost of the induction equation) for axisymmetric flows in cylindrical coordinates
$(r,\varphi,z)$ \citep{lgj06}. The implementation of fully insulating and partially conducting boundary conditions are discussed in \citet{lgj07}. 
The computational domain is shown in Fig.~\ref{princeton_full_scheme} 
and the parameters are summarized in Table~\ref{Ta_parameter}.
  Six coils with dimensions as shown were used,
  with $67$ turns in the two coils nearest the midplane and $72$ in the rest. They are split into two sets of three in parallel, with the upper three in series and the bottom three also in series.
  Currents $I_\varphi$ were adjusted according to the experimental values. Note that in the simulations the magnetic diffusivity $\eta_{\rm Ga}$ is fixed to the experimental value $\eta_{\rm Ga}=2,430\;\unit{cm^{2}\;s^{-1}}$ (Table.\ref{Ta_parameter}), however the kinematic viscosity is varied for the purpose of extrapolation from numerically
tractable to experimentally realistic Reynolds numbers.

\begin{figure}[!htp]
\subfigure{\scalebox{0.4}{\includegraphics{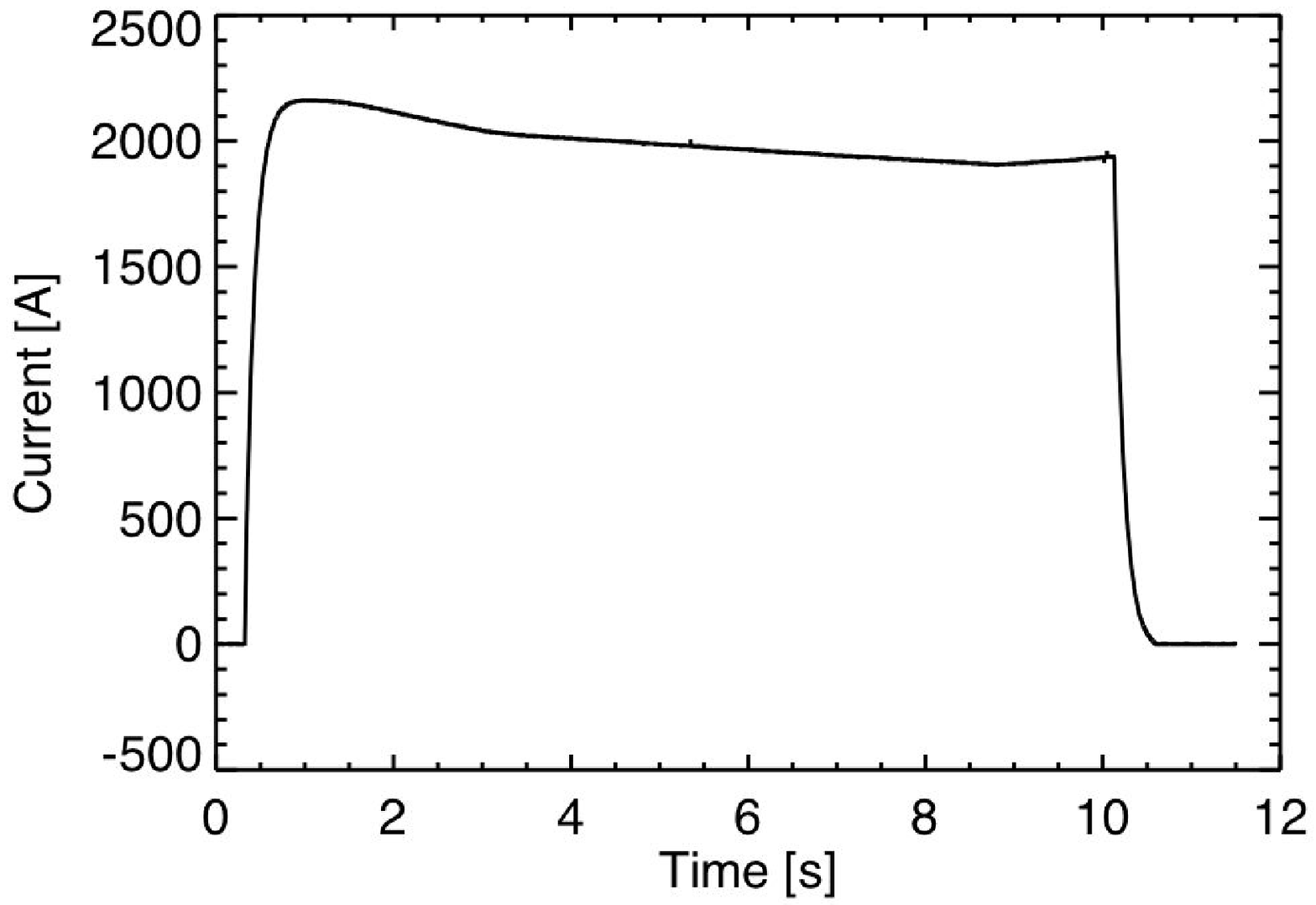}}}$\;$
\subfigure{\scalebox{0.4}{\includegraphics{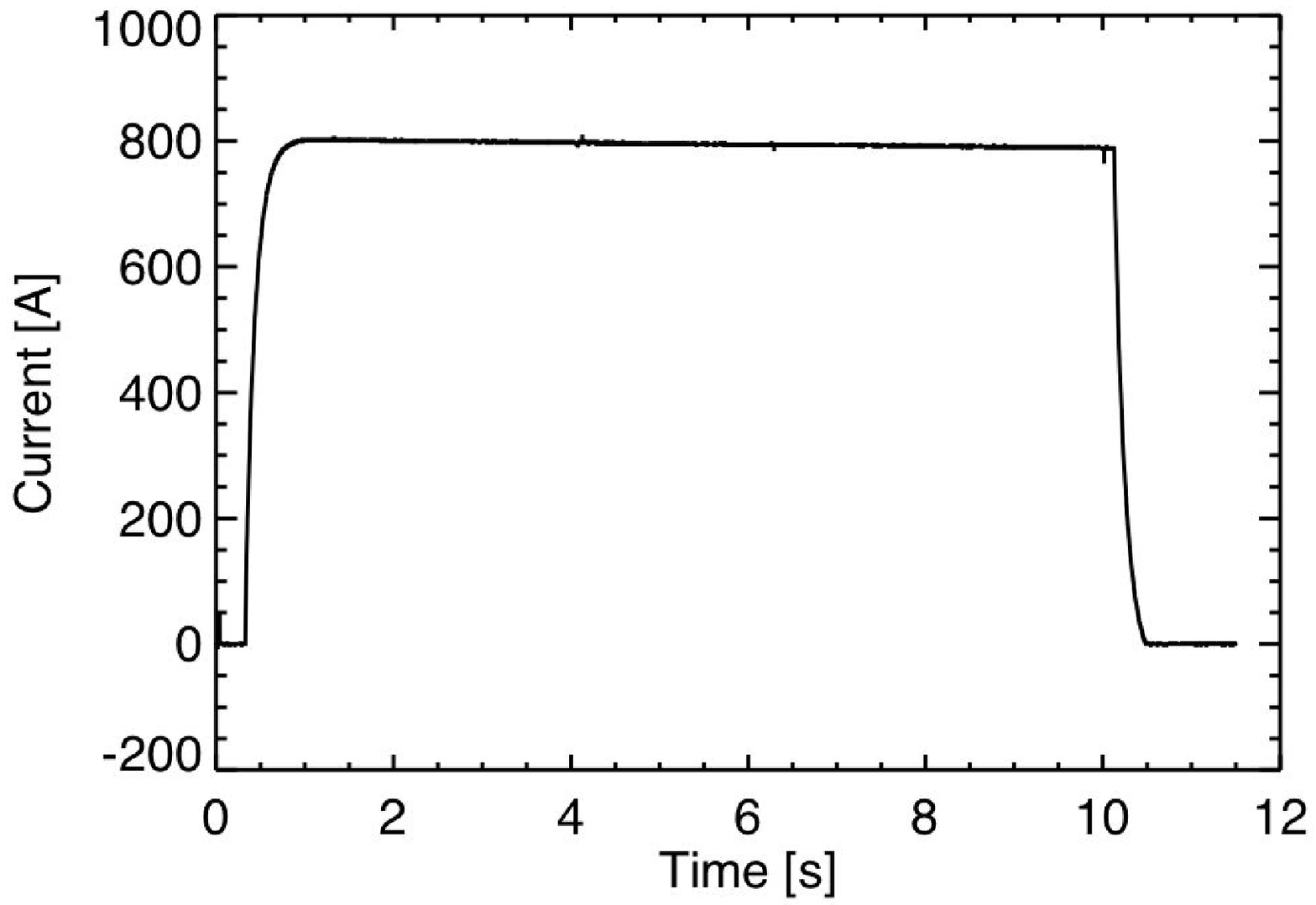}}}
\caption{\label{experiment_coil} Figure Courtesy of Mark Nornberg. Experimental wave form of the coil current. Left panel: $I_{\varphi}=1000\;\unit{A}$; right: $I_{\varphi}=400\;\unit{A}$. Shown here is the total current in the general circuit. Since the six coils are split into two sets of three in parallel, with the upper three in series and the bottom three also in series, the current in the coils should be the one in the general circuit divided by $2$. }  
\end{figure}

\begin{figure}[!htp]
\begin{center}
\scalebox{0.6}{\includegraphics{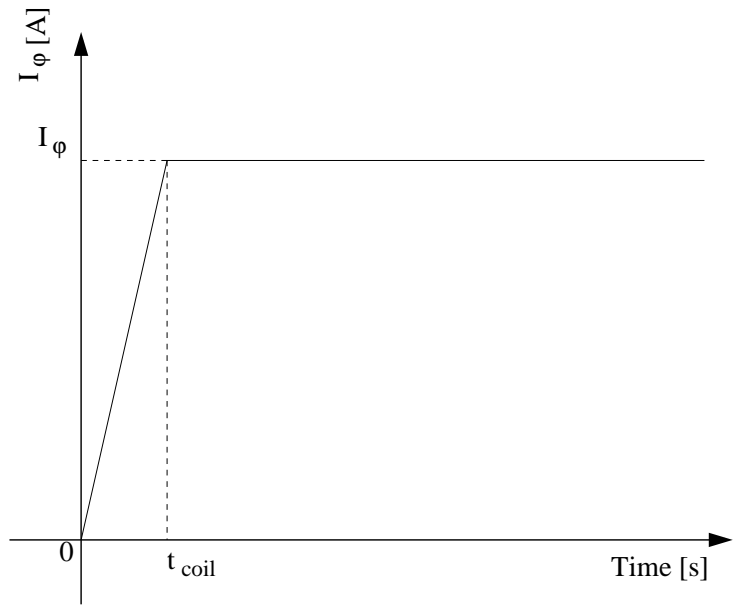}}
\caption{\label{coil} Simulated Current pattern in the external coils; the ramp time $t_{coil}=0.2\;\unit{s}$; except where stated explicitly, the
steady current $I_{\varphi}=1000\;\unit{A}$.}
\end{center}
\end{figure}

The measured current waveform is displayed in Fig.~\ref{experiment_coil} for target
currents of $1000\;\unit{A}$ and $400\;\unit{A}$. The waveform displayed has an overshoot at the early stage, a linear
decline and then a linear ramp where the controller tries to adjust the output voltage to reach the programmed set point. This behavior is peculiar to the high
current runs (left panel of Fig.~\ref{experiment_coil}). Lower currents have a much flatter waveform (right panel of Fig.~\ref{experiment_coil}). In the simulation a wave form like that in Fig.~\ref{coil} is used to approximate the experimental coil currents with ramp time $t_{\rm coil}=0.2\;\unit{s}$.

Since the container is made of stainless steel, not a perfectly
insulator, the radial magnetic angular momentum flux ($-rB_\varphi
B_r/4\pi$) at the cylinders need not vanish.  In principle, the
magnetic coupling of the fluid to the cylinders might modify the
growth of the instability\citep{lgj07}.

Fortunately, at the frequencies relevant to Princeton MRI experiment (100\% run, $\bar{\Omega}=\sqrt{\Omega_1\Omega_2}\sim150\unit{Hz}$) 
the skin depth of stainless steel $\delta_{w}=\sqrt{2/\sigma_{\rm Steel}\bar{\Omega}\mu_{0}}\approx9\unit{cm}$ ($\sigma_{\rm Steel}$ is the conductivity of stainless steel, see the value in Table.~\ref{Ta_parameter}), which is much larger than the
thickness of the steel vessel surrounding the gallium in the
experiment, $d_{w}\approx1.0\unit{cm}$, so that the magnetic field
diffuses rather easily into the boundary.  
%\remark{Say $\eta_{\rm steel}$
%$\eta_{\rm Fe}$ since the steel is not pure iron. Quote the value of
%$\eta_{\rm steel}$ here or in the Table. I gather from the ratio of resistances
%stated below that $\eta_{\rm steel}\approx 2.3\eta_{\rm Ga}$ (?)} 
If one
considers axial currents, the gallium and the steel wall act as
resistors in parallel; taking into account their conductivities and
radial thicknesses, one finds that the resistances of steel walls are much larger than the resistance of the liquid gallium
[$R_{I}:R_{II}:R_{III}=65:1:21$; see Fig.~\ref{princeton_full_scheme} for the
subscripts].  Therefore, the currents carried by the steel walls
can be neglected for the toroidal field, so that the $B_r B_\varphi$ stress at
the boundary is expected to be unimportant and an insulating boundary condition
suffices.

The linear growth rate and saturated final state based on partially conducting boundary condition differs slightly from the results based on perfectly insulating boundary condition with partially conducting walls, which verifies our argument. Hereafter we ignore the conductivity of the stainless steel walls, but regard them as insulating materials, that is: $\eta_{I}=\eta_{II}=\infty$. Please note that in this whole paper the rotation profile used is $[\Omega_1,\Omega_2,\Omega_3,\Omega_4]=[4000,533,1820,650]\;{\rm rpm}$ (100\% run, $\Omega_4>\Omega_2$, $\Omega_1$, $\Omega_2$, $\Omega_3$, and $\Omega_4$ are the rotation speed of the inner cylincer, outer cylinder, inner ring and outer ring, respectively) (Table.\ref{Ta_parameter}), under which the simulations predict that the bulk flow is almost ideal Couette state \citep{kjg04}. And this is the desired basic state for MRI to grow. However, the experimental scans find that only if we choose $[\Omega_1,\Omega_2,\Omega_3,\Omega_4]=[400,53,146,40]\;{\rm rpm}$ (10\% run, $\Omega_4<\Omega_2$), the bulk flow profile is found to be very close to the ideal Couette state with the deviation (the boundary layer) highly confined near both endcaps \citep{jbsg06}. \citet{jbsg06} also shows that non-magnetic quasi-keplerian flows at Reynolds numbers up to millions are essentially steady, if the boundary effects have been properly considered. The Stewartson layer at the junction of the rings may be smoothed by localized circulation and/or turbulence from these instabilities \citep{lw08}. The rotation profile to be used in the magnetized runs would be $[\Omega_1,\Omega_2,\Omega_3,\Omega_4]=[2400,320,876,240]\;{\rm rpm}$ (60\% run, $\Omega_4<\Omega_2$). The difference of the rotation profile at the end caps between in the experiment and in the simulation could possibly be explained by the wobbling of the inner cylinder in the experimental runs due to the difficulties of aligning the inner cylinder perfectly and the gaps between the rings and cylinders \citep{rjlg07}. And these effects are speculated to be more important with higher Reynolds number $Re\sim10^6-10^7$ as in the experiment. Unfortunately, the modern computers and codes can not afford a simulation with Reynolds number as high as several million. Although this is not experimentally realizable, the low-Reynolds-number results reported in this paper without wobbling, gaps and starting from an initially ideal Couette state with fully insulating magnetic boundary conditions still gives us some useful hints about what is going on in the experiment. The result maybe is even closer to what is happening in the experiment based on the purely hydrodynamic experiment (steady flow, mostly ideal Couette state in the bulk), although this argument needs to be proven. We are addressing the problem step by step, adding the complexity into the code one by one. To include all subtle points at one time would perplex the problem and make it difficult for us to catch the essence of the problem. The detailed study of the influence of the wobbling and gaps between the rings is on the way to investigate the discrepancies between the experimental results and simulations. Also we want to point out:  as demonstrated by \citet{gj02}, the viscosity of liquid metals is so small as to be almost irrelevant to MRI, at least in the linear regime. Thus a low-Reynolds-number run would be fine, at least in this sense.

Most of results presented in this paper are the results of 100\% run except stated explicitly. The simulation results have proved very valuable in design, operation and understanding of the experiment. The outline of this paper is as follows.  We present the results of the linear MRI in Sec.~\ref{linear}. And the results of the nonlinear saturation of MRI are discussed in Sec.~\ref{nonlinear}. Sec.~\ref{summary} summarizes the results and presents the final conclusions. 

\section{Linear Phase}\label{linear}

\subsection{Linear Growth Rate Reduced by the Residual Magnetized Ekman Circulation}

\begin{figure}[!htp]
\begin{center}

\subfigure{\scalebox{0.4}{\includegraphics{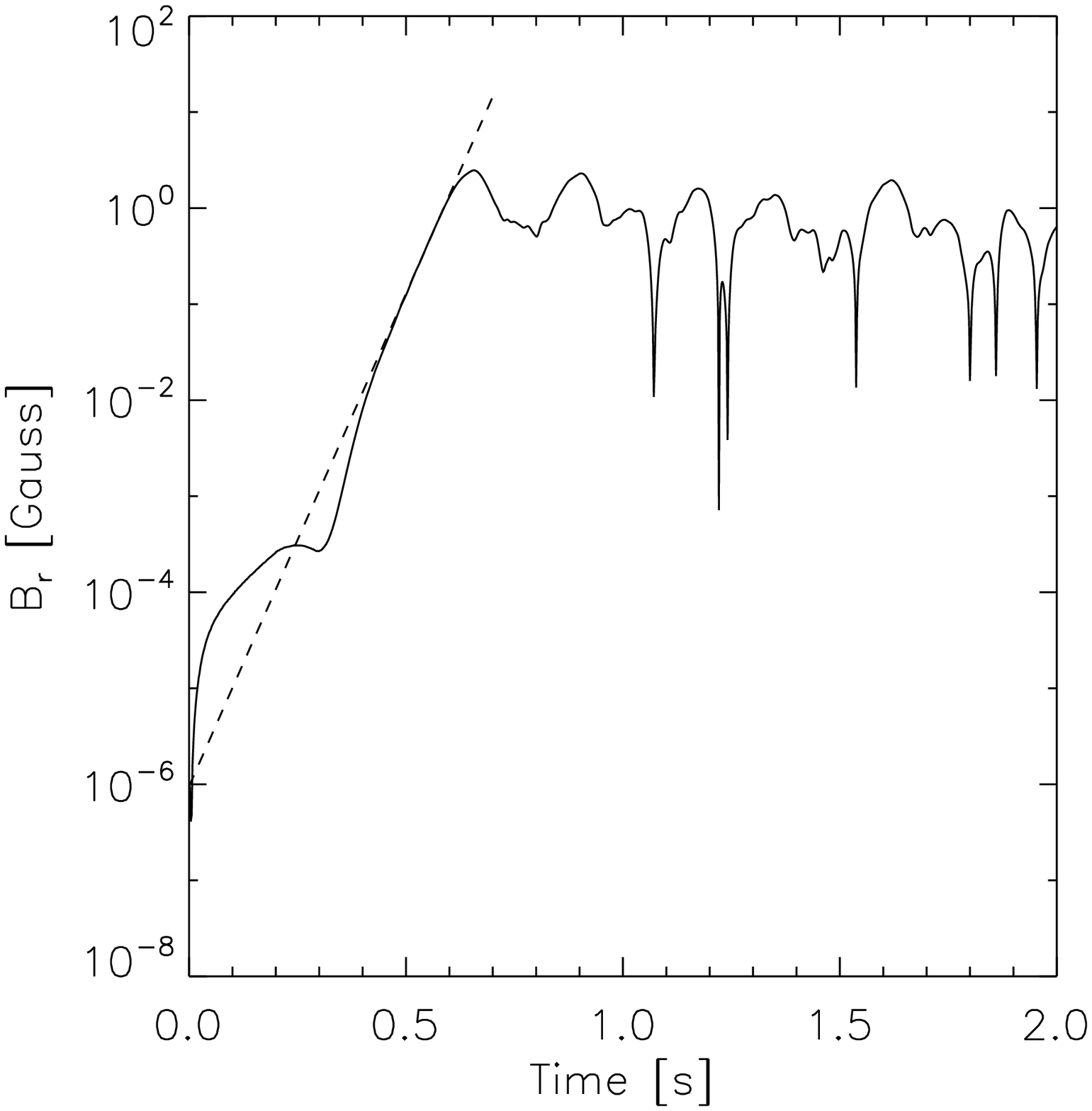}}}$\,$
\subfigure{\scalebox{0.4}{\includegraphics{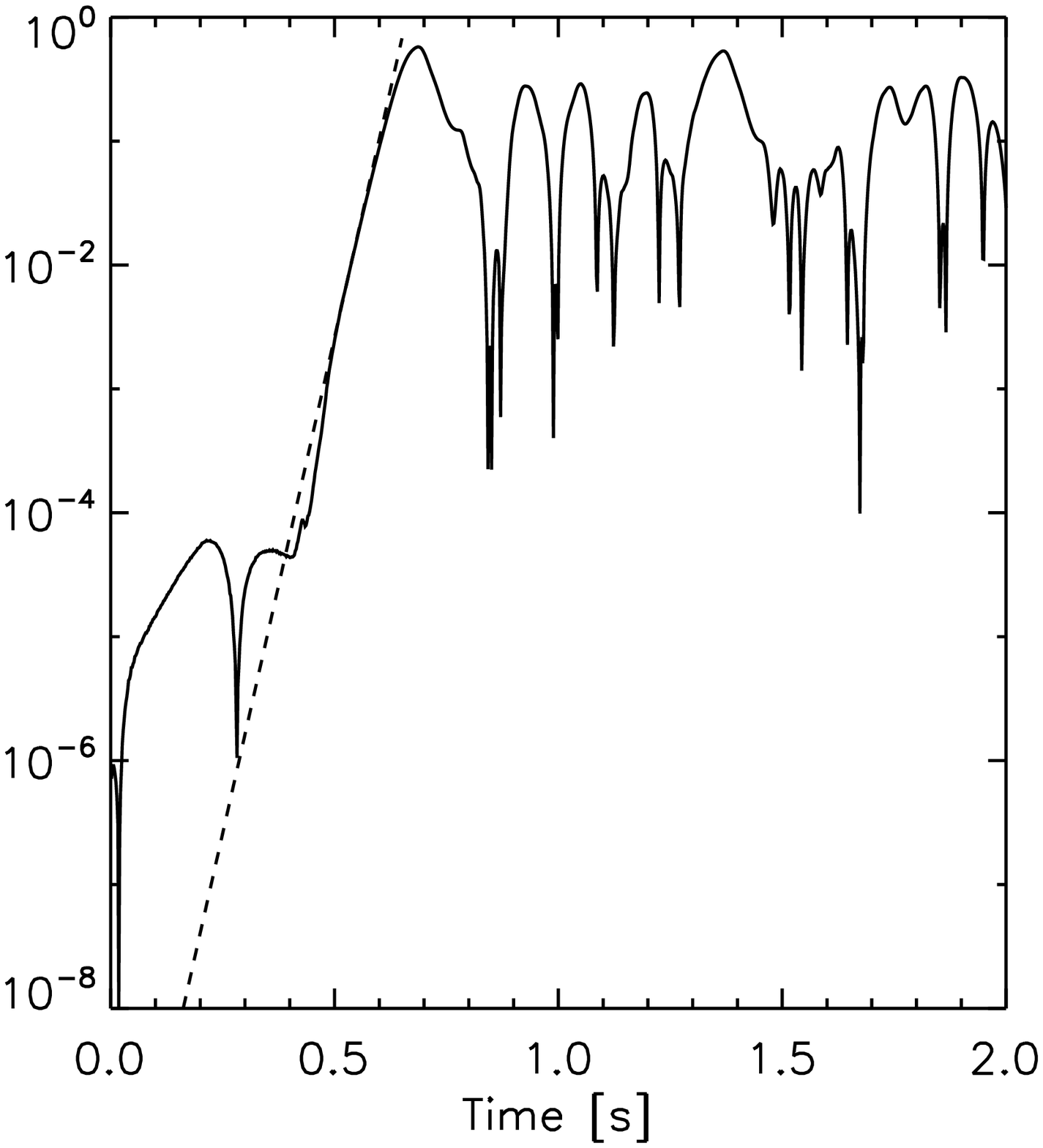}}}
\caption{\label{Princeton_Ring2_Br_outside} 100\% run (MRI unstable) and $I_{\varphi}=1000\;\unit{A}$. $B_{r}$ \emph{vs.} time for $Re=6400$, $\Rm=20$ sampled outside the fluid at $z=13.95\;\unit{cm}$, $r=25.0\;\unit{cm}$.``Bottom end cap'' is located at $z=0$. Height $h=27.9\;\unit{cm}$. Left panel: endcaps split into two rings, Growth rate $\gamma=21.7\;\unit{s^{-1}}$; right panel: ideal Couette state at both endcaps, Growth rate $\gamma=33.1\;\unit{s^{-1}}$.}
\end{center}
\end{figure}

The first convincing evidence of the existence of MRI is its linear growth rate (Fig.~\ref{Princeton_Ring2_Br_outside}). We find this linear growth rate is reduced from $33.1\;\unit{s^{-1}}$ with ideal Couette state at both endcaps (left panel of Fig.~\ref{Princeton_Ring2_Br_outside}, the end effect is removed completely by enforcing ideal Couette state at both endcaps, thus no Ekman circulation is present) to $21.7\;\unit{s^{-1}}$ with two rings just as in the real experiment (right panel of Fig.~\ref{Princeton_Ring2_Br_outside}). This is due to the residual magnetic Ekman circulation, which modifies the background flow. Also as pointed out in \citet{lw08b}, the linear growth rate of an absolute instability in a bounded geometry is reduced by the ``absorping" end plate, with the reduction extent proportional to $O(\Gamma^{-2})$, where the aspect ratio is $\Gamma=h/(r_2-r_1)$ and $h$ is the height of the cylinders (see Fig.~\ref{princeton_full_scheme}). 
%\remark{Don't you mean $O(\Gamma^{-2})$ here?}

Interestingly the linear growth rate with ideal Couette state at both endcaps (right panel of Fig.~\ref{Princeton_Ring2_Br_outside}) matches the growth rate of the fastest growing mode, which naturally is the dominant mode, got from a linear code developed by \citep{gj02}, in which vertical periodicity is assumed, but the radial equations are
solved directly by finite differences with perfectly insulating
boundary conditions (Table~\ref{Princeton_Compare_with_Global}).  
The agreement suggests that viscous effects are slight, since the no-slip conditions
on the horizontal velocities at the endcaps differ from the periodicity imposed by
the linear code.

\subsection{Strong Magnetic Field Suppresses MRI with Two Split Rings}

\begin{figure}[!htp]
\subfigure{\scalebox{0.4}{\includegraphics{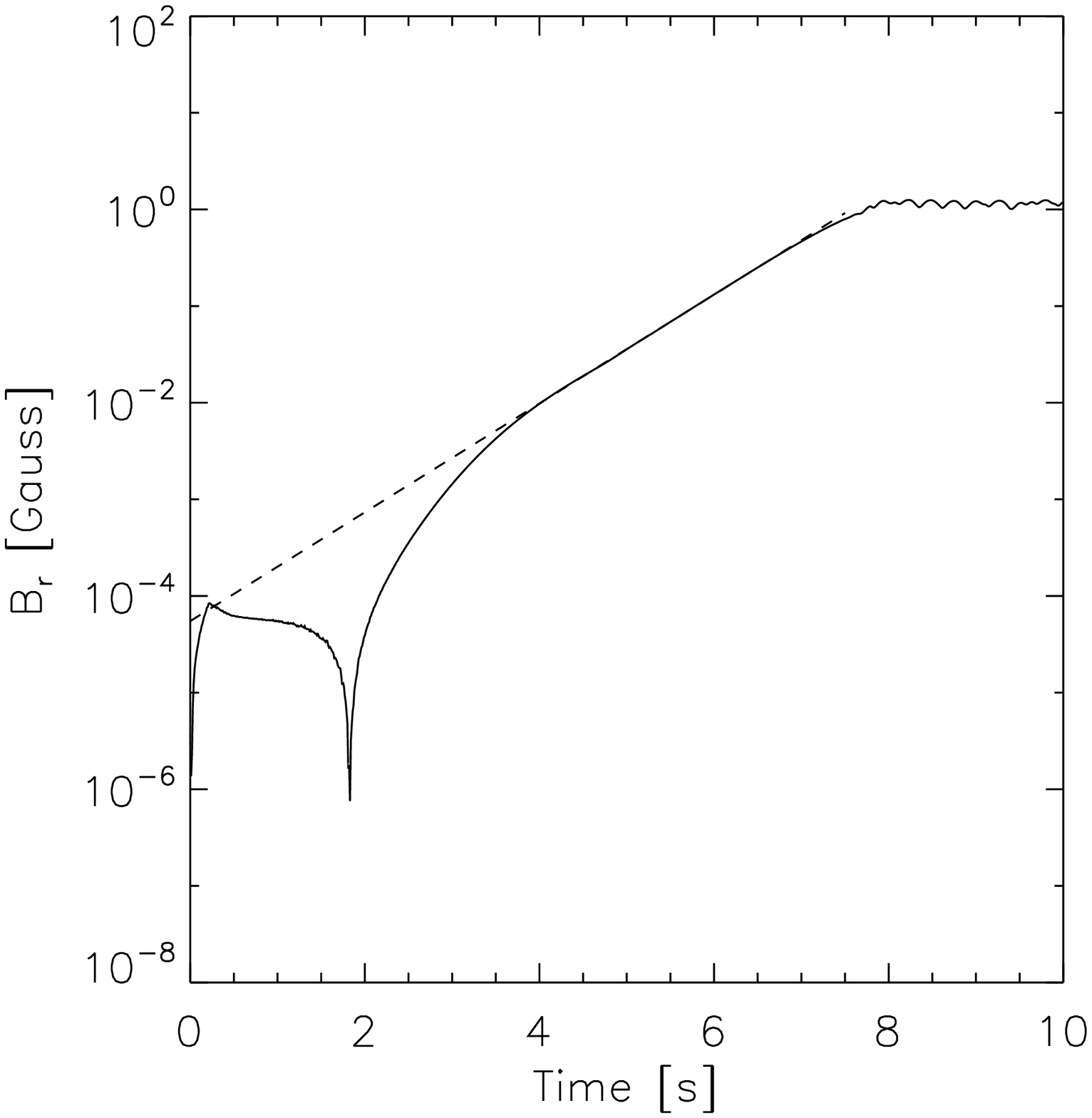}}}$\,$
\subfigure{\scalebox{0.4}{\includegraphics{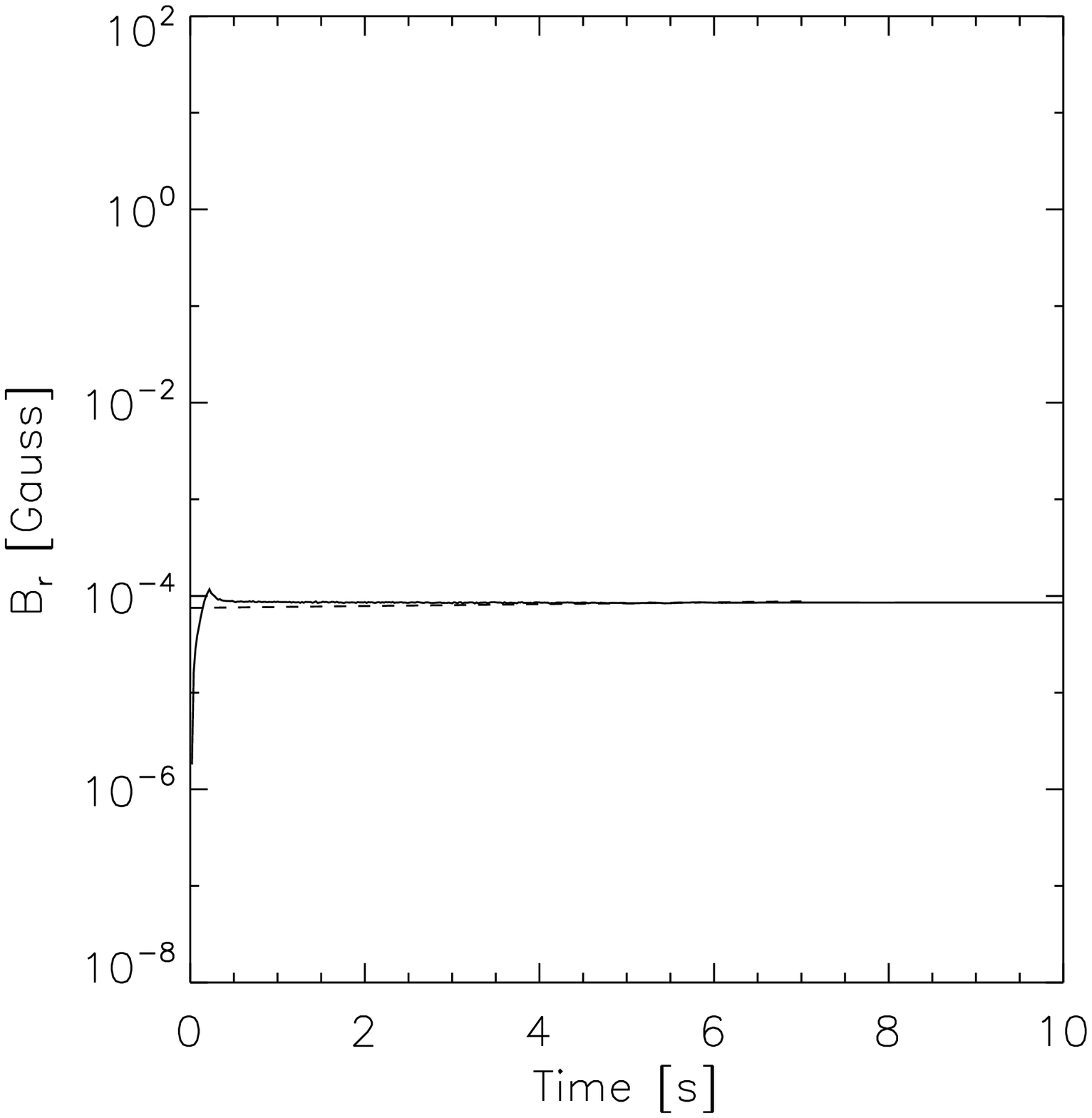}}}
\caption{\label{suppress} 45\% run. $B_{r}$ \emph{vs.} time for $Re=6400$, $\Rm=7.3$ sampled outside at $z=13.95\;\unit{cm}$, $r=25.0\;\unit{cm}$.``Bottom end cap'' is located at $z=0$. Height $h=27.9\;\unit{cm}$. left panel: $I_{\varphi}=750\;\unit{A}$, growth rate $\gamma=1.3\;\unit{s^{-1}}$; right panel: $I_{\varphi}=1200\;\unit{A}$, stable.}   
\end{figure}

MRI essentially is a weak field instability. It is characteristic of MRI for a strong magnetic field to suppress this instability. Our experimental facility can only allow $I_{\varphi}\lesssim1200\;\unit{A}$. Thus we need to try carefully to find one set of proper parameters under which one growing mode is present with low $I_{\varphi}$ while absent with larger $I_{\varphi}$. 
Fig.~\ref{suppress} demonstrates this property for
 simulations in which the boundary
rotation rates $(\Omega_1,\Omega_2,\Omega_3,\Omega_4)$ are scaled
to 45\% of their designed values (Table.\ref{Ta_parameter}).

\section{Nonlinear Saturation}\label{nonlinear}

\subsection{Inflowing ``Jet" Observed near the Middle Plane}\label{inflowing_jet}

\begin{figure}

\scalebox{0.4}{\includegraphics{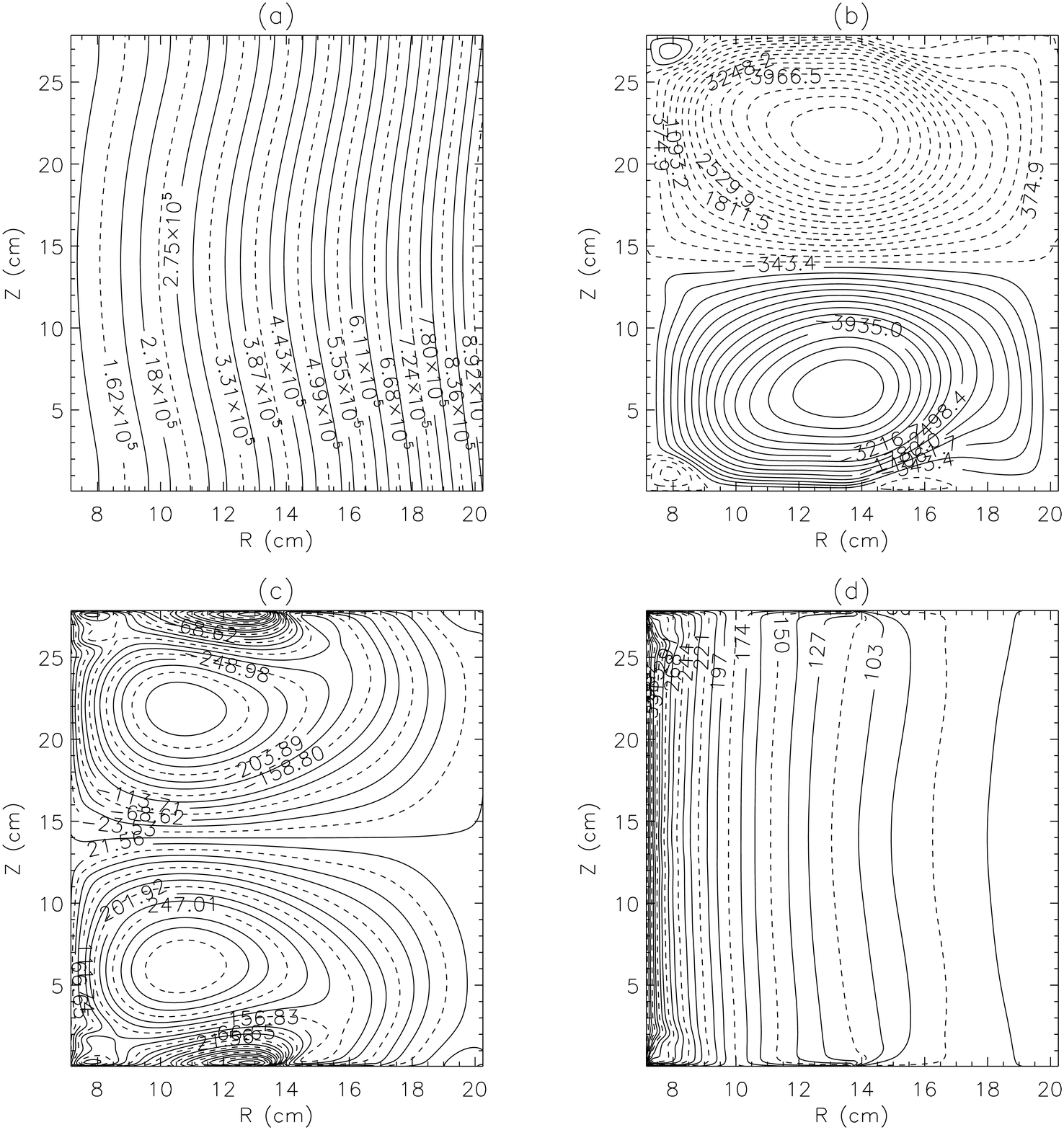}}
\caption{\label{Princeton_Ring2_final_pattern} 100\% run. Contour plots of final-state
velocities and fields (MRI unstable).  $Re=6400$, $\Rm=20$. $I_{\varphi}=1000\;\unit{A}$, $\Lambda=0.36$. (a) Poloidal flux
function $\Psi \unit{(Gauss\,cm^{2})}$ (b) Poloidal stream function
$\Phi \unit{(cm^{2}s^{-1})}$ (c) toroidal field
$B_{\varphi} \unit{(Gauss)}$ (d) angular velocity
$\Omega\equiv r^{-1}V_{\varphi}\unit{(rad\,s^{-1})}$. In panel (b), dashed line indicates the clock-wise poloidal circulation while solid line indicates the anti-clock-wise poloidal circulation.}
\end{figure}

For $Re=6400$, 
the final state is not steady.  Typical time averaged flow and
field patterns are shown in
Fig.~\ref{Princeton_Ring2_final_pattern}.  The poloidal flux
and stream functions are defined so that
\begin{equation}\label{chap3_pfuncs}
\boldsymbol{V}_P\equiv
V_r\boldsymbol{e}_r+V_z\boldsymbol{e}_z=r^{-1}\boldsymbol{e}_\varphi
\boldsymbol{\times\nabla}\Phi,\qquad \boldsymbol{B}_P\equiv
B_r\boldsymbol{e}_r+B_z\boldsymbol{e}_z=r^{-1}\boldsymbol{e}_\varphi
\boldsymbol{\times\nabla}\Psi,
\end{equation}
which imply $\boldsymbol{\nabla\cdot V}_P=0$
and $\boldsymbol{\nabla\cdot B}_P=0$.  The divergence of the velocity field
in these compressible but subsonic simulations is nonzero but small.
%[Our velocity field is slightly compressible, so that eq.~(\ref{chap3_pfuncs})
%does not quite capture the full velocity field.
%Nevertheless, the error is small, and 
%$\Phi$ is well defined by
%$\nabla^2(\Phi\boldsymbol{e}_\varphi/r)=
%\boldsymbol{\nabla\times V}_{\rm P}$ with $\partial\Phi/\partial r=0$ on the end caps and $\partial\Phi/\partial z=0$ on the cylinders.]

\begin{figure}
\begin{center}

\scalebox{0.4}{\includegraphics{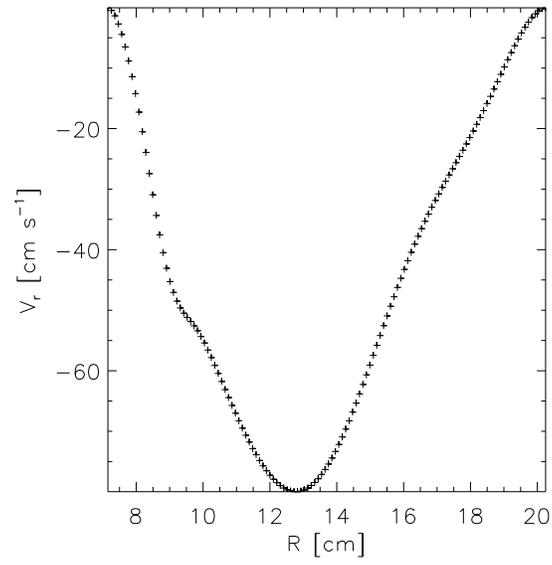}}
\caption{\label{Ring2_100_1000_vr} Corresponding to Fig.~\ref{Princeton_Ring2_final_pattern}(b). 100\% run (MRI unstable). Time averaged $v_{r}$ \emph{vs.} radius $r$ on the midplane $(z=h/2)$. $Re=6400$, $\Rm=20$. $I_{\varphi}=1000\;\unit{A}$, $\Lambda=0.36$.}
\end{center}
\end{figure}

We note that the induced toroidal field is around $6\%$ of the initial axially imposed magnetic field at this magnetic Reynolds number $\Rm$:
$B_{\varphi,\max}\approx 0.06 B_{\infty}$. 
The most striking
feature is the inflowing ``jet'' centered near $z=13.95\;\unit{cm}$
in Fig.~\ref{Princeton_Ring2_final_pattern} (see also Fig.~\ref{Ring2_100_1000_vr}), which is opposite to the 
usual Ekman circulation \citep{kjg04}. It seems that the rapid outflowing ``jet" found in \citep{lgj06} with vertically periodic boundary condition, where the position of the ``jet" is arbitrary, is shifted to the boundary layer, near the end caps in the experiment. 

The inward ``jet" near the inner cylinder part is a direct consequence of MRI. 
For MRI-stable regimes, \citet{lw08} discuss the influence of
independently controlled split endcaps upon the poloidal circulation
and the influence of the axial magnetic field and Reynolds number upon
the Stewartson and Ekman layers.
In such regimes, an \emph{outflowing} jet occurs near the inner cylinder,
in our apparatus, as it does in purely hydrodynamic Ekman circulation (i.e.,
$\Lambda=0$).
 However in the MRI-unstable regime for the same Elsasser number $\Lambda$, the middle clock-wise cell enlarges both horizontally and vertically and eventually dominates the other cells, which retreat to the corners if the instability is strong enough.  This leads to one inflowing ``jet" near the inner cylinder (Fig.~\ref{Ring2_100_1000_vr}). 

\begin{figure}[!htp]
\begin{center}

\scalebox{0.4}{\includegraphics{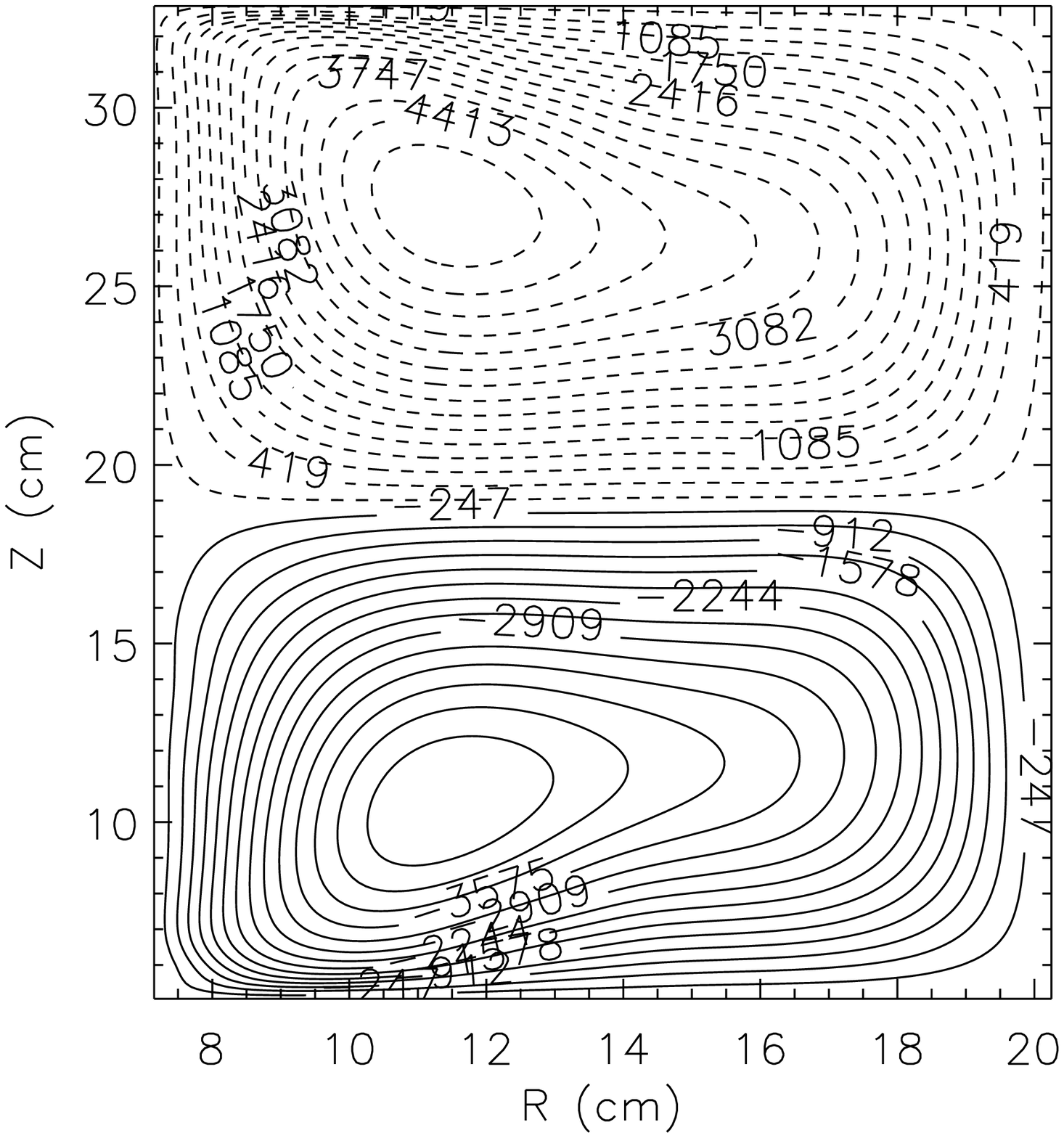}}
\caption{\label{Princeton_Couette_stream} 100\% rotation run (MRI unstable). Contour plots of
 poloidal stream function $\Phi \unit{(cm^{2}s^{-1})}$ for $Re=6400$ with ideal Couette state at the end caps. $I_{\varphi}=1000\;\unit{A}$, $\Lambda=0.36$. Dashed line indicates the clock-wise poloidal circulation while solid line indicates the anti-clock-wise poloidal circulation.}
\end{center}
\end{figure}

\begin{figure}[!htp]
\begin{center}

\scalebox{0.4}{\includegraphics{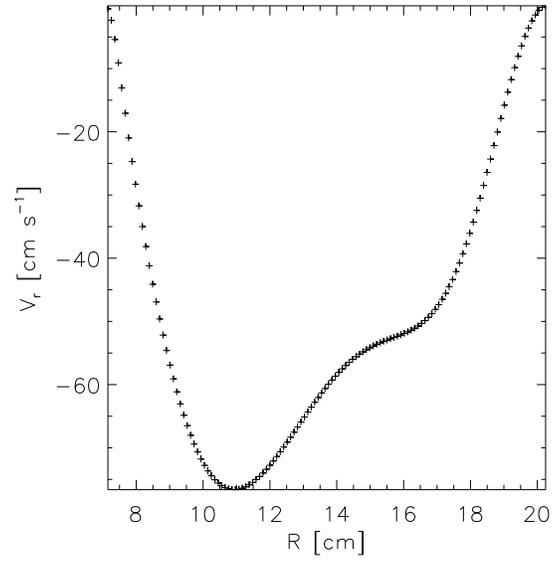}}
\caption{\label{Ideal_100_1000_vr} Corresponding to Fig.~\ref{Princeton_Couette_stream}. 100\% rotation run (MRI unstable). Time averaged $v_{r}$ \emph{vs.} radius $r$ on the midplane $(z=h/2)$ for $Re=6400$ with ideal Couette state at the end caps. $I_{\varphi}=1000\;\unit{A}$, $\Lambda=0.36$.}
\end{center}
\end{figure}

Further evidence that MRI causes the inflowing jet was found
in simulations with the ideal Couette state imposed at the end caps, to remove
Ekman circulation  (Fig.~\ref{Princeton_Couette_stream}). 
%From Fig.~\ref{Princeton_Couette_stream},  the direction (clockwise) of circulation 
%is opposite to the usual Ekman circulation (anticlockwise) \citep{kjg04}. 
Fig.~\ref{Princeton_Couette_stream} shows a clear inflowing jet near the inner cylinder in this case (see also Fig.~\ref{Ideal_100_1000_vr}). This suggests that the poloidal circulation seen in the final state in the split-endcap cases (Fig.~\ref{Princeton_Ring2_final_pattern} (b)) is caused solely by saturation of MRI rather than 
(magnetically modified) Ekman circulation.

\subsection{MRI Enhances Radially Outward Angular Momentum Transport at Saturation}\label{angular_ideal}

\begin{figure}[!htp]
\begin{center}

\subfigure{\scalebox{0.4}{\includegraphics{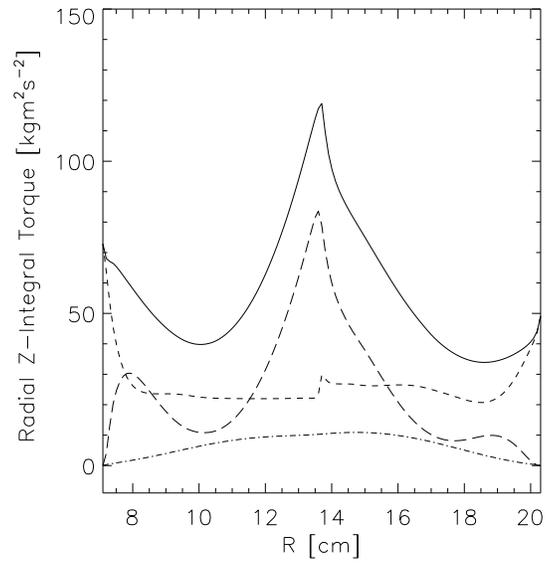}}}
\caption{\label{Princeton_Ring2_integral_torque_r} 100\% run (MRI unstable). Time averaged $z$-integrated radial angular momentum fluxes versus $r$ at saturation.
$Re=6400$, $\Rm=20$. $I_{\varphi}=1000\;\unit{A}$, $\Lambda=0.36$. Dash line~, viscous torque; dash dot line~, magnetic torque; long dash line~, advective torque; solid line, total torque. }
\end{center}   
\end{figure}

Astrophysicists are interested in the angular momentum transport due to MRI since MRI is supposed to be the most probable mechanism to explain the fast accretion in the astrophysical disks.  Fig.~\ref{Princeton_Ring2_integral_torque_r} displays
the time-averaged $r$-profiles of the radial advective, viscous, and magnetic torques, i.e. the
angular momentum fluxes integrated over cylinders coaxial with the boundaries:
\begin{eqnarray}
\Gamma_{\rm advective,r}(r)&=&2\pi\int_{0}^{h} dz\,\rho r^{2} v_{r}v_{\varphi}\, ,\\
\Gamma_{\rm magnetic,r}(r)&=&2\pi
\int_{0}^{h} dz \left(-\frac{r^{2}B_{r}B_{\varphi}}{4\pi}\right)\, ,\\
\Gamma_{\rm viscous,r}(r)&=&2\pi\int_{0}^{h} dz 
\left[-r^{3}\rho \nu \frac{\partial}{\partial r}
\left(\frac{v_{\varphi}}{r}\right)\right]\\[1ex]
\Gamma_{\rm total,r}(r)&=&
\Gamma_{\rm advective,r}(r)+\Gamma_{\rm magnetic,r}(r)+\Gamma_{\rm viscous,r}(r)\, ,
\end{eqnarray}
where $dr$ and $dz$ are the radial and vertical cell sizes respectively.

\begin{figure}[!htp]
\begin{center}

\subfigure{\scalebox{0.4}{\includegraphics{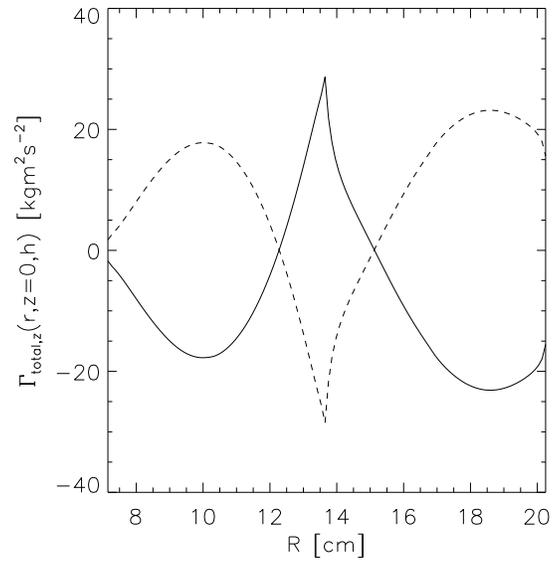}}}
\caption{\label{Princeton_Ring2_integral_torque_z} 100\% run (MRI unstable). Time averaged vertical total angular momentum flux at both endcaps versus $r$ at saturation.
$Re=6400$, $\Rm=20$. $I_{\varphi}=1000\;\unit{A}$, $\Lambda=0.36$. Solid line: bottom endcap $(z=0)$; dashed line: top endcap $(z=h)$.}

\end{center}   
\end{figure}

In contrast to the final state for vertically periodic boundary
conditions \citep{lgj06}, the total radial torque is not constant with
radius.  Since our numerical scheme conserves angular momentum
exactly, we can infer a vertical flux arising from
exchange of angular momentum with the endcaps (see also Fig.~\ref{Princeton_Ring2_integral_torque_z}).  In the statistical steady state (nonlinearly saturated final state) the sum of the angular momentum flux on all boundaries should be zero, at least on time average. In this case it is fluctuating since the final state is not steady given high Reynolds number. 
From the gradients of
the radial torque, we identify four Ekman
circulation cells: where $d\Gamma_{\rm total,r}/dr>0$ ($<0$) , the fluid is losing
(gaining) angular momentum at the endcaps and the boundary-layer flow
is therefore radially inward (outward).  This is consistent with the
discussion in \citet{lw08} of the poloidal circulation driven by two
split rings.  The radial magnetic and advective torques vanish at $r_1$ and $r_2$ because
of the boundary conditions but are important at intermediate radii, 
especially the advective.
All components of the radial torque are positive, which means that the angular momentum is transported radially outwards.

Fig.~\ref{Princeton_Ring2_integral_torque_z} displays
the time-averaged $r$-distributions of the vertical total
angular momentum flux at both endcaps $(z=0,h)$:
\begin{eqnarray}
\Gamma_{\rm advective,z}(r,z=0,h)&=&2\pi \int_{r_1}^{r} dr \rho r^{2} v_{z}v_{\varphi}\, ,\\
\Gamma_{\rm magnetic,z}(r,z=0,h)&=&2\pi \int_{r_1}^{r} dr \left(-\frac{r^{2}B_{z}B_{\varphi}}{4\pi}\right)\, ,\\
\Gamma_{\rm viscous,z}(r,z=0,h)&=&2\pi \int_{r_1}^{r} dr 
\left[-r^{2}\rho \nu \frac{\partial}{\partial z}
\left(v_{\varphi}\right)\right]\,.%\\[1ex]
%\Gamma_{\rm total,z}(r,z=0,h)&=&
%\Gamma_{\rm advective,z}(r,z=0,h)+\Gamma_{\rm magnetic,z}(r,z=0,h)+\Gamma_{\rm viscous,z}(r,z=0,h)\, .
\end{eqnarray}
%where $\Delta r$ is the radial cell size in the simulations. 
Since at both endcaps, both the advective and magnetic angular momentum fluxes are zero due to the boundary conditions, the total angular momentum at both endcaps $\Gamma_{\rm total,z}(r,z=0,h)=
\Gamma_{\rm advective,z}(r,z=0,h)+\Gamma_{\rm magnetic,z}(r,z=0,h)+\Gamma_{\rm viscous,z}(r,z=0,h)$ are simply the viscous angular momentum flux $\Gamma_{{\rm viscous},z}(r,z=0,h)$.

From these two figures, we can see that the global angular momentum is entering from the inner cylinder then most of it is flowing out from the outer cylinder while the rest of it is flowing out at the two end caps. 
\emph{Angular momentum is transported radially outward.}

\begin{figure}
\scalebox{0.4}{\includegraphics{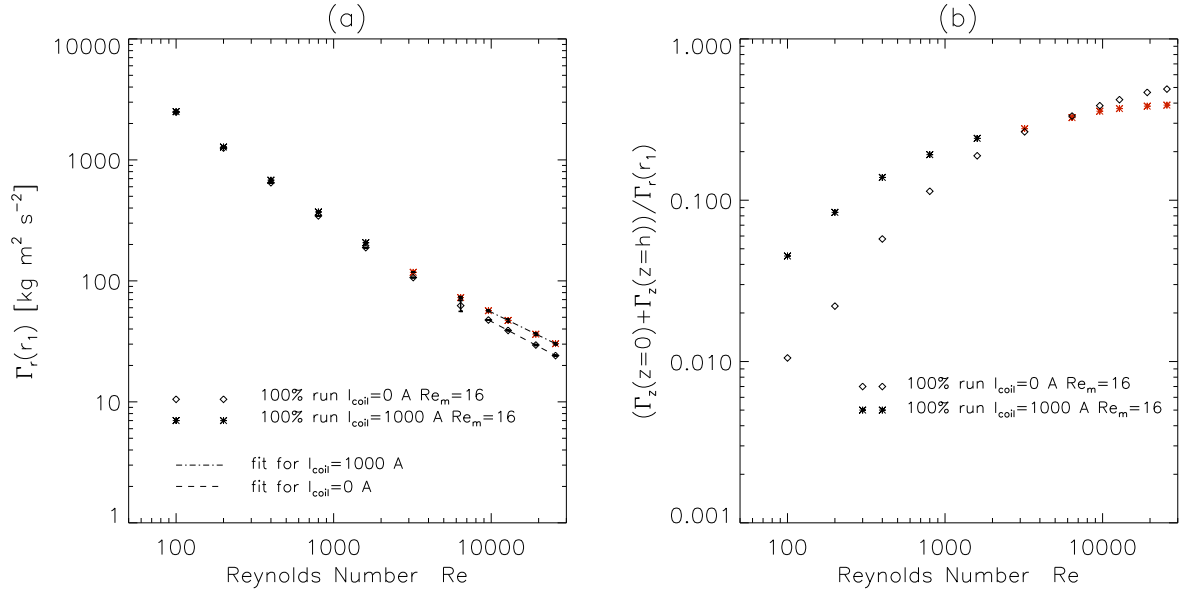}}
\caption{~(Color) (a) Total radial torque at the inner cylinder (b) Sum of the total vertical torques at both endcaps \emph{versus} $Re$. Note that in the simulations the magnetic diffusivity $\eta$ is fixed to the experimental value $\eta=2430\;\unit{cm^{2}\;s^{-1}}$ (Table.\ref{Ta_parameter}), however the kinetic viscosity is varied for the purpose of extrapolation.
In both panel, red colors: MRI unstable; black color: MRI stable. In panel (a), dashed lines have slopes of $-0.691$ (initial state) and
dash-dot lines $-0.639$ (final state). The error bars on panel (a) (very small, typically $\sim1\%$, hard to see on panel (a)) are the standard time-deviations. If $Re<3200$, the error bars are zero in both magnetized and unmagnetized cases, which suggests the system is steady with low Reynolds number.
}
\label{100_ratio}

\end{figure}

It is very interesting to derive the dependence of the total torque at the inner cylinder, which is mainly responsible for driving the rotation, 
and of the sum of the vertical torques at both endcaps, on Reynolds number ($Re$).
From Fig.~\ref{100_ratio} (a), we infer the following scalings ($100\%$ run, $\Rm=16$):

(1) In the absence of magnetic field (100\% run, $100\lesssim Re\lesssim25600$),
\begin{equation}\label{torque_scaling1}
\Gamma_{{\rm initial}, r}(r_{1})\approx 2.69\times10^{4}Re^{-0.691}\, ;
\end{equation}
These hydrodynamic states (rather than ideal Couette flow) will be
used as the initial conditions for the MRI experiments. The maximum relative error ratio, defined as the ratio of the absolute difference of the value from the simulation and the fitted value (Eq.~\ref{torque_scaling1}) over the former one, is $\lesssim0.4\%$.

(2) For $I_{\varphi}=1000\;\unit{A}$ and $\Lambda=0.36$ (100\% run, $100\lesssim Re\lesssim25600$). 
\begin{equation}\label{torque_scaling2}
\Gamma_{{\rm final},r}(r_{1})\approx 1.98\times10^{4}Re^{-0.639}\, .
\end{equation}
This corresponds to the final states of the MRI experiment. The maximum relative error ratio is $\lesssim0.3\%$.
From these two scaling laws, \emph{the MRI indeed enhances the angular momentum transport at saturation, though slightly} (see discussion below). The enhancement is far beyond the arrange of the error bar (see the error bars on panel (a) of Fig.~\ref{100_ratio}). It is reasonable for $\Gamma_{r}$ in both magnetized and unmagnetized cases to decrease with increasing $Re$ since the viscous coupling to the walls scales as $Re^{-1/2}$ in standard unmagnetized Ekman layers (\emph{i.e.} the endcaps are corotating solidly with the outer cylinder). In this more complicated case (with split rings), we also expect  
a similar relationship, though the magnetized Ekman and Stewartson layers complicate the problem \citep{lw08}.

If Eq.~\ref{torque_scaling1} and Eq.~\ref{torque_scaling2} also work at larger Reynolds numbers, so that they may safely be extrapolated to the experimental Reynolds number ($Re\approx1.15\times10^{7}$, 100\% run), then the total radial torque of the initial and final state at the inner cylinder may be as large as
\begin{equation}
\Gamma_{{\rm initial},r}(r_{1})\approx 0.359\;\unit{[Newton\; m]}\nonumber
\end{equation}
and 
\begin{equation}
\Gamma_{{\rm final},r}(r_{1})\approx 0.611\;\unit{[Newton\; m]}\nonumber
\end{equation}
respectively. Thus, the ratio of the increase of the torque over the initial torque is: $(0.611-0.356)/0.356=72\%$, which is quite measurable and indicates that at the experimental Reynolds number MRI would dominate the residual magnetic Ekman circulation in the point of view of transporting the angular momentum. There are, however, reasons for caution in accepting this estimate. For example, 
the experimental flow may be three-dimensional and
turbulent, which might result in an even higher torque in the final state, and the absolute values of both the exponents seem to decrease at larger Reynolds number and the difference of these two exponents is small. These concerns all make the extrapolation of Eq.~\ref{torque_scaling1} and Eq.~\ref{torque_scaling2} to the experimental Reynolds number  a bit risky. 
Nevertheless, we expect a noticeable torque enhancement in the
MRI-unstable regime. 

From Fig.~\ref{100_ratio} (b), we can see that:
(1) At larger Reynolds number, $(\Gamma_{z}(z=0)+\Gamma_{z}(z=h))/\Gamma_{r}(r=r_1)$ is increased, which means that larger part of the total angular momentum is transported vertically. 
%This is reasonable since the Ekman circulation is enhanced when Reynolds number is larger;
(2) In the MRI stable regime ($Re\lesssim1600$), the magnetic field enhances the vertical transport of the angular momentum. This is also reasonable since the magnetic field would align the flow, thus having the cells elongating and penetrating deeper into the bulk. The size of the middle cells is increased vertically by the residual magnetic Ekman circulation. These are consistent with the conclusions deduced in \citet{lw08}.
(3) In the MRI unstable regime ($Re\gtrsim3200$), the onset of the MRI results in more angular momentum transported radially outwards and less vertically. The MRI would increase the scale of the middle cell horizontally. Therefore it transports more angular momentum radially outwards.

\section{Conclusions}\label{summary}

In conclusion of this paper we have simulated the nonlinear development
of magnetorotational instability in a nonideal magnetohydrodynamic
Taylor-Couette flow.  The simulations mimic an on-going experiment 
except that the conductivity of the stainless steel walls is neglected and the simulation is started from an ideal Couette state rather than an actual hydrodynamic
statistical steady state 
%\remark{this clumsy phrase is more accurate
%than ``equilibrium,'' since there are fluctuations and dissipation}
driven by split end caps. We
have also restricted our study to smaller fluid Reynolds number ($Re$) than in the
experiment, however we have used exactly the same magnetic Reynolds number ($\Rm$). MRI grows from small amplitudes at rates in good agreement
with linear analyses without the end cap effects.

Concerning the MRI simulations with two split independently rotating rings like the real experimental facility, we draw the following conclusions:
\begin{enumerate}
\item A clear linear phase is observed; the linear MRI growth rate is reduced by the residual magnetized Ekman circulation.
\item Strong magnetic field suppresses MRI.
\item In the final state one inflowing ``jet" opposite to the usual Ekman circulation ``jet" \citep{kjg04} is found near the inner cylinder, a direct consequence of MRI rather than the residual Magnetic Ekman circulation (100\% run).
\item The MRI enhances the angular momentum transport at saturation. (100\% run).
\item The final state contains horizontal fields about $6\%$ as large as the initial vertical field for $\Rm\approx 20$ (100\% run).
\end{enumerate}

We emphasize that these conclusions are based on axisymmetric
simulations restricted to the range $10^2\lesssim Re,\Rm\lesssim 10^{4.4}$ with the idealizations mentioned above. The simulation results considering the conductivity of the steel container, starting from an actual hydrodynamical equilibrium and the comparison with the experimental results would be given in a future paper.

%\begin{acknowledgements}
The author would like to sincerely thank Jeremy Goodman and Hantao Ji for their very inspiring discussion and constructive comments. The author would also like to thank James Stone for the advice on the ZEUS
code and thank Stephen Jardin for the advice of implementing full insulating boundary condition. This work was supported by the US Department of Energy, NASA
under grants ATP03-0084-0106 and APRA04-0000-0152 and also by the
National Science Foundation under grant AST-0205903.
%\end{acknowledgements}

%\bibliographystyle{apj} %% or your favorite style

%\bibliography{thesis}

\clearpage

 \begin{table}
 \begin{center}
 \begin{tabular}{| c | c |} \hline
 \multicolumn{2}{| c |}{ Dimensions} \\ \hline
 $r_{1}=7.1\;\unit{cm}$ & $r_{2}=20.3\;\unit{cm}$ \\ \hline
 $h=27.9\;\unit{cm}$     & $d_{w}=0.9525\;\unit{cm}$ \\ \hline
 \multicolumn{2}{| c |}{Material Properties} \\ \hline 
 $\rho_{\rm{Ga}}=6.35 \;\unit{g\;cm^{-3}}$ & $\eta_{\rm{Ga}}=2.43\times10^{3}\;\unit{cm^{2}\;s^{-1}}$ \\ 
 \hline
%  \multicolumn{2}{| c |}{$\eta_{\rm{Steel}}=5.73\times10^{3}\;\unit{cm^{2} s^{-1}}$ } \\ \hline
 $\eta_{\rm{Steel}}=5.73\times10^{3}\;\unit{cm^{2}\;s^{-1}}$ & $\sigma_{\rm Steel}=1.25\times10^{16}\;\unit{s^{-1}}$ \\ 
 \hline  
   \multicolumn{2}{| c |}{ Rotation Profile (100\% run)} \\ \hline
 $\Omega_{1}/2\pi=4000\;\unit{rpm}$ & $\Omega_{2}/2\pi=533\;\unit{rpm}$ \\ \hline
 $\Omega_{3}/2\pi=1820\;\unit{rpm}$ & $\Omega_{4}/2\pi=650\;\unit{rpm}$ \\ \hline
 \multicolumn{2}{| c |}{ Rotation Profile (45\% run)} \\ \hline
 $\Omega_{1}/2\pi=1600\;\unit{rpm}$ & $\Omega_{2}/2\pi=239.85\;\unit{rpm}$ \\ \hline
 $\Omega_{3}/2\pi=819\;\unit{rpm}$ & $\Omega_{4}/2\pi=292.5\;\unit{rpm}$ \\ \hline 
 \end{tabular}
  \caption{Parameters in Gaussian units used in the simulations}\label{Ta_parameter}

 \end{center}
 \end{table}

\begin{table}
\begin{center}

\begin{tabular}{|c|c|c|c|c|}
\hline 
$\Rm$&
$Re$&
$~n~$ &
Prediction $[\unit{s}^{-1}]$&
Simulation $[\unit{s}^{-1}]$ \tabularnewline
\hline\hline
16&
6400&
1&
33.7&
33.1
\tabularnewline
16&
6400&
2&
13.8&
\tabularnewline
\hline 
\end{tabular}
\caption{\label{Princeton_Compare_with_Global} 100\% run. Growth rates from
semianalytic linear analysis \emph{vs.} simulation.}
\end{center}
\end{table}

\end{document}